\documentclass{article}
\usepackage{amsmath,amsfonts,amssymb}
\newtheorem{theorem}{Theorem}

\begin{document}

\title{A correct proof of the heuristic GCD algorithm.}\author{Bernard
Parisse\\
Institut Fourier\\
Université de Grenoble I}\maketitle

\begin{abstract}
  In this note, we fill a gap in the proof of the heuristic GCD in the
  multivariate case made by Char, Geddes and Gonnet ({\cite{gcdheu}}) and give
  some additionnal information on this method.
\end{abstract}

\section{Context}

The heuristic gcd algorithm is used to computed the gcd of two polynomials $P$
and $Q$ with integer coefficients in one or a few variables : the main idea is
to evaluate one of the variable $X_k$ at a sufficient large integer $z$, 
compute the gcd of the evaluations recursively or as integers and 
reconstruct a candidate gcd from the gcd of the evaluations using the
representation of coefficients in basis $z$ with symmetric representation. It
was introduced 15 years ago and is used intensively in popular CAS like Maple
or MuPAD, see \cite{liaoevaluation}
for more details on the efficiency of this algorithm. 

The proof given in the paper of Char, Geddes and Gonnet is correct
in one dimension but is wrong in the multivariate case. Indeed, in the proof
of lemma 2 (p.37), the authors applies the univariate case demonstration to a
polynomial they call $P^{( 1 )}$ at a point $\alpha$ that fullfills the
hypothesis (6) of lemma 2 for the polynomial $P$, but they don't check that
$\alpha$ fullfills this hypothesis (6) for the polynomial $P^{( 1 )}$. And
there is no reason for $\alpha$ to fullfill it since $P^{( 1 )}$ is obtained
by evaluation of all but one variable at integers that sometimes must be
non-zero or might even be very large (the keypoint for the evaluation point of
the other variables is that the main coefficient of $P$ with respect to the
$X_k$ variable does not evaluate to zero which implies that the main
coefficient of $Q$ also does not evaluate to zero). Correcting the lemma with
the same proof would require for example that hypothesis (6) would be replaced
by :
\[ | \alpha | \geq 1 + | P^{( 1 )} | \]
where $P^{( 1 )}$ can be any evaluation of $P$ for the variables $X_j \neq
X_k$ at integers so that the degree of $P^{( 1 )}$ is the same as the degree
of $P$ with respect to $X_k$. This has three problems :

\begin{itemize}
  \item it can increase the size of $\alpha$ (which will reduce the efficiency
  of the algorithm)
  
  \item it requires an additional step of evaluation of a polynomial at
  non-zero values.
  
  \item it would require fixing the code in CAS using it
\end{itemize}

Fortunately, we will give an alternative proof of the correctness of the
algorithm as it is implemented in most CAS, extend the ring to the Gaussian
integers, and give another lower bound for the evaluation point that insures
we get the gcd (this lower bound has probably only a theoretical interest)

\begin{theorem}
  Let $P$ and $Q$ be two polynomials depending on the variables $X_1, \ldots
  ., X_k$, with integer coefficients or with Gaussian integer coefficients. We
  use the notation :
  \[ P ( z ) := P ( X_1, \ldots ., X_{k - 1}, z ) \]
  Let $z$ be any integer such that $| z | \geq 2 \ast \min ( | P |, | Q |
  ) + 2$, where $| P |$ denotes the largest norm of all the coefficients of
  $P$. Assume that the primitive part $G$ of the $z$-adic symmetric
  reconstruction of $\gcd ( P ( z ), Q ( z ))$ divides both $P$ and $Q$.
  
  Then $G$ is the gcd of $P$ and $Q$. The assumption that $G$ divides
  $P$ and $Q$ is always true for $z$ sufficiently large.
\end{theorem}

\section{Proof of theorem 1.}

Let $g = \gcd ( P ( z ), Q ( z ))$. From the definition of $G$, if $\alpha$ is
the integer content of the $z$-adic symmetric reconstruction of $g$, we have :
\begin{equation}
  g = \alpha G ( z ), | \alpha | \leq \frac{| z |}{2} \label{g1}
\end{equation}
If $D$ is the polynomial gcd of $P$ and $Q$, then $D ( z )$ divides both $P (
z )$ and $Q ( z )$, therefore
\begin{equation}
  g = \beta D ( z ), \beta \in \mathbb{Z}[X_1,...,X_{k-1}] \label{g2}
\end{equation}
If $G$ divides $P$ and $Q$, $G$ divides $D$, hence there exists a polynomial
$C$ such that~:
\begin{equation}
  D = C G \Rightarrow  D ( z ) = C ( z ) G ( z ) \label{g3}
\end{equation}
Combining (\ref{g1}), (\ref{g2}) and (\ref{g3}), we get :
\begin{equation}
  \alpha G ( z ) = \beta C ( z ) G ( z )
\end{equation}
We want to prove that $C$ is a constant polynomial. We have the relation
\begin{equation}
  \text{$\alpha = \beta C ( z )$ where $\beta \in \mathbb{Z},$} | \alpha |
  \leq \frac{| z |}{2} \label{cz}
\end{equation}
Therefore $C ( z )$ does not depend on the variables $X_1, \ldots ., X_{k -
1}$. More precisely, there is a polynomial $C'$ with integer coefficients such
that :
\begin{equation}
  C = C ( z ) + ( X_k - z ) C'
\end{equation}
We want to prove that $C'$ is zero. Assume that $C' \neq 0$.

We begin by showing that $C'$ can not depend on $X_1, \ldots ., X_{k - 1}$
(this is the new multivariate step). Indeed, if $C'$ has degree $d_1 \neq 0$
with respect to $X_1$ for example, then the highest degree term of $C'$ with
respect to $X_1$ is $c_1 ( X_2, \ldots ., X_k ) X_1^{d_1}$, hence the highest
degree term of $C$ with respect to $X_1$ is $( X_k - z ) c_1 X_1^{d_1}$. Since
$C$ divides both $P$ and $Q$, this highest degree term $( X_k - z ) c_1
X_1^{d_1}$ divides the highest degree term $p_1 X_1^{d_{P, 1}}$ of $P$ and
$q_1 X_1^{d_{Q, 1}}$ of $Q$ with respect to $X_1$. Therefore $X_k - z$ divides
$p_1 ( X_2, \ldots ., X_k )$ and $q_1 ( X_2, \ldots ., X_k )$. Now we look at
the lowest non-zero degree term of $p_1$ and $q_1$ with respect to $X_k$:
these polynomials of the variables $X_2, \ldots ., X_{k - 1}$ are divisible by
$z$. Since they are not zero, we conclude that at least one non-zero
coefficient of $P$ and $Q$ is divisible by $z$. This is a contradiction to the
hypothesis $| z | \geq 2 \ast \min ( | P |, | Q | ) + 2$.

We are now reduced to prove the unidimensionnal case since 
$C$ depends only on the
variable $X_k$ and the proof of the original article applies, for the sake of
completness, let us recall briefly this proof (see also \cite{gonnet} for a
proof in dimension 1). The idea is to factor $C$ over
$\mathbb{C}$ :
\begin{equation}
   C ( X_k ) = c_k \prod_{j = 1}^{\mbox{degree} ( C )} ( X_k - z_j )
     \label{factor} 
\end{equation}
Since $C$ divides $P$, $C ( X_k )$ divides $P ( 0, \ldots ., 0, X_k )$,
therefore the roots $z_j$ of $C$ are also roots of $P ( 0, \ldots ., 0, X_k
)$. Same for $Q$. Therefore, there exists a subset of coefficients of $P$ or
of $Q$, therefore bounded by $\min ( | P |, | Q | )$, such that
\begin{equation} 
   \sum_{l = 0}^m a_l x^l = 0 \label{root}, \mbox{ for } x = z_j 
\end{equation}
It is well known that (\ref{root}) implies :
\begin{equation} \label{broot}
  | x | < \frac{A}{| a_m|} + 1, \quad A=\max_{0 \leq i \leq m - 1}( | a_i | )
\end{equation}
Indeed, if $|x| \leq 1$, (\ref{broot}) is trivial (because $A=0$ implies
$x=0$). Otherwise~:
\[|a_m x^m|=|-\sum_{i=0}^{m-1} a_i x^i| \leq A \sum_{i=0}^{m-1} |x|^i
=A\frac{|x|^m-1}{|x|-1}\]
therefore, since $|x|-1 > 0$~:
\[ |a_m| |x|^m (|x|-1) \leq A |x|^m < A |x|^m\]
which implies (\ref{broot}). 

Now equation (\ref{broot}) 
gives the bound $| z_j | < | z | / 2$ for all $j$. Applying this bound
to (\ref{factor}), we get :
\[ | C ( z ) | \geq | c_k | \prod_{j = 1}^{\mbox{degree} ( C )} ( | z | -
   | z_j | )  > \left( \frac{| z |}{2} \right)^{\mbox{degree} (
   C )} \]
which contradicts (\ref{cz}). This ends the proof that $G$ is the gcd of $P$
and $Q$.

Note that during the whole proof, we can replace the coefficient ring
$\mathbb{Z}$ by $\mathbb{Z} [ i ]$ without any changes~: the gcdheu algorithm
works if the coefficients are Gaussian integers.

We finish by giving a theoretical lower bound on $z$ such that $G$ will always
divide $P$ and $Q$. This bound will involve the extended gcd
algorithm (Bézout identity) on $P$ and $Q$. Let us assume first than
we are in dimension 1. Since $\gcd ( P, Q ) = D$, there
exists an integer $\gamma$ and polynomials $U$ and $V$ with integer
coefficients such that :
\begin{equation}
  P U + Q V = \gamma D
\end{equation}
At the point $z$, we get that $\gamma D ( z )$ is in the ideal $< P ( z ), Q (
z ) > = < g >$, hence $g$ divides $\gamma D$$( z )$. We already know from
(\ref{g2}) that $g = \beta$$D ( z )$ where $\beta$ is an integer in
the univariate case. Therefore $\beta$ divides $\gamma$. Now assume that 
\begin{equation} \label{estimate}
 | z| > 2 | D | | \gamma |
\end{equation}
where the lower bound depends only of the original
polynomials $P$ and $Q$. If this assumption is fullfilled, then $| z | > 2 | D
| | \beta |$ and the symmetric $z$-adic representation of $g = \beta D ( z )$
is the polynomial $\beta D$. The primitive part of $\beta D$ is $D$, hence $G
= D$.
In dimension greater than one, trying to apply the same idea
will work but with a small modification. Indeed $\beta$ and
$\gamma$ are now polynomials of the variables $X_1,...,X_{k-1}$.
To conclude, we have two choices~:
\begin{itemize}
\item we accept a denominator depending on $X_1,..,X_{k-1}$ during the division
test of $P$ and $Q$ by $G$. In this case, the lower bound  (\ref{estimate})
on $|z|$ should be $2 |D|$ times the Landau-Mignotte bound on coefficients 
of the factors of $P$ and $Q$,
\item we remove the gcd of the coefficients
of $P$ and $Q$ viewed as polynomials in $X_k$ with coefficients in
$\mathbb{Z}[X_1,...,X_{k-1}]$. Then $\beta$ is an integer dividing the
polynomial $\gamma$ and the lower bound (\ref{estimate}) is correct.
\end{itemize}

\bibliographystyle{abbrv}
\bibliography{gcdheu}

\end{document}